\documentclass[twocolumn,showpacs,a4paper]{revtex4} 

\usepackage{amsmath} 
\usepackage{bm}

\begin{document} 

\title{Reduction of the RPA eigenvalue problem 
and a generalized Cholesky decomposition 
for real-symmetric matrices}

\author{P.~Papakonstantinou} 
\email[Email:]{panagiota.papakonstantinou@physik.tu-darmstadt.de} 
\affiliation{Institut f\"ur Kernphysik, 
Technische Universit\"at Darmstadt, 
Schlossgartenstr.~9, 
D-64289 Darmstadt, Germany}

\begin{abstract} 
The particular symmetry of the random-phase-approximation (RPA) 
matrix has been utilized in the past 
to reduce the RPA eigenvalue 
problem into a symmetric-matrix problem of half the dimension. 
The condition of positive definiteness of at least one of the matrices $A\pm B$ has been imposed 
(where $A$ and $B$ are the submatrices of the RPA matrix) so that, {\em e.g.}, 
its square root can be found by 
Cholesky decomposition. 
In this work, alternative methods are pointed out 
to reduce the RPA problem to a real (not symmetric, in general) problem 
of half the dimension, 
with the condition of positive definiteness relaxed.  
One of the methods relies on a generalized Cholesky decomposition, 
valid for non-singular real symmetric matrices. 
The algorithm is described and a corresponding routine in C is given.
\end{abstract}


\maketitle

\section{Introduction} 
\label{Sintr} 

The eigenvalue problem of 
random-phase-approximation (RPA) type, 
\begin{equation} 
  \left( 
  \begin{array}{rr} 
       A &  B \\ 
      -B & -A  
  \end{array} 
  \right) 
  \left( 
  \begin{array}{c} 
       X_{\lambda} \\ Y_{\lambda} 
  \end{array} 
  \right) 
\equiv 
{\bf R} 
  \left( 
  \begin{array}{c} 
       X_{\lambda} \\ Y_{\lambda} 
  \end{array} 
  \right) 
= \varepsilon_{\lambda} 
\left( 
  \begin{array}{c} 
       X_{\lambda} \\ Y_{\lambda} 
  \end{array} 
\right) ,
\label{Erpa}  
\end{equation} 
where  $A$ and $B$ are real symmetric $n\times n$ matrices, 
is frequently encountered in quantum many-body physics. 
In nuclear physics, in particular, the RPA is the most widely used method 
to examine collective excitations of nuclei. 
The RPA is also used to compute the correlation energy of a system described at zeroth order 
by the Hartree-Fock approximation. 
Standard first-order RPA, relativistic RPA, renormalized RPA, as well as 
quasi-particle and second-order RPA, when formulated in configuration space, result in equations of the form (\ref{Erpa}). 
The properties of the solutions, following from the symmetry of the matrix $\bf R$, 
are well known~\cite{Tho1960,Tho1961,Row1970,RiS1980}. 

The $(2n)\times (2n)$ RPA matrix $\bf R$ is not symmetric --- 
one reason why for large sizes $n$ it becomes prohibiting to solve the RPA problem as-is. 
It has been possible, however, 
to exploit the particular structure of $\bf R$, determined by the symmetric matrices $A$ and $B$, in order  
to reduce the RPA problem not only to a symmetric eigenvalue problem, but 
at the same time a problem of half the dimension, $n\times n$~\cite{Chi1970,UlR1971}. 
(The generalized RPA problem where $A$ is Hermitian and $B$ symmetric 
can also be reduced~\cite{UlG1972}, but that problem will not be dealt with here.) 
As long as one wishes to avoid complex matrices, 
that can be achieved under the condition that at least one of the matrices $A+B$, $A-B$ 
be positive definite. 
The Cholesky decomposition of the positive-definite matrix 
$A\pm B$ lies at the heart of the method presented in Ref.~\cite{Chi1970}, 
while in Ref.~\cite{UlR1971} an orthogonal transformation is utilized 
under the same condition.  

It can happen, however, that the 
positive-definiteness condition of $A\pm B$ is not met, 
meaning that the RPA matrix 
has imaginary eigenvalues or the stability matrix is not positive-definite. 
In modern relativistic RPA models of nuclear response the $A\pm B$ matrices are known to be indefinite, 
due to the inclusion of states from the Dirac sea~\cite{Rin2001}. 
Thus the full $(2n)\times (2n)$ non-symmetric problem is currently solved when using 
relativistic RPA in configuration space.  
Other situations from nuclear physics include the trivial case of a dipole spurious state appearing at imaginary energy, 
as well as systems which are unstable against 
certain ``excitations". 
An exotic nucleus with different enough proton- and neutron-Fermi energies 
can be found unstable against 
configurations of isospin $T=1$ and certain angular momentum and parity $J^{\pi}$. 
Also in the neighborgood of phase transitions the stability matrix can have negative 
eigenvalues when the residual interaction 
is attractive and large enough. 
In such cases the (in)definiteness of the matrices may not be known 
before solving the RPA problem.  

In this work alternative procedures are proposed for reducing the size of the RPA problem, 
still involving real matrices, where the condition 
of positive-definitness is relaxed. 
The result is a real, non-symmetric (in general) problem of half the original dimension.  
One procedure requires no matrix decomposition and is 
recommended for problems where the $A\pm B$ matrix is expected from the 
outset to not be positive-definite.  
Another method offers the possibility to 
detect and solve a symmetric problem if the matrix turns out to be positive-definite. 
The latter involves a more general Cholesky-like decomposition of a real 
symmetric matrix. 
The way to perform the decomposition in practice is outlined
and 
a C routine is also provided. 
Compared with the usual Cholesky decomposition, the additional computational 
and storage effort is minimal.  
Other reduction methods can be devised based on different matrix 
decompositions, as will be demonstrated. 

Modern computers perform quite well in solving the 
standard first-order RPA in nuclei, without reduction.  
Still, the need to reduce the RPA problem and thus accelerate its solution 
becomes rather pressing when, 
{\em e.g.}, 
one solves the 
RPA equations iteratively (as in renormalized RPA) 
or one wants to evaluate the RPA correlation 
energy of a nucleus. In such cases many RPA problems corresponding to different 
$J^{\pi}T$ quantum numbers must be solved (perhaps more than once) before a solution is reached. 
Hence a fast technique is of great importance. 

Reduction methods are of interest also when one needs to evaluate only the 
lowest positive eigenvalues of very large RPA matrices, using, {\em e.g.}, the 
Lanczos technique -- see Ref.~\cite{Tsi2001} for a related strategy. 

%
Next it is  
briefly reviewed how the RPA problem can be reduced 
using the Cholesky decomposition. 
Then 
alternative methods  
and the modified Cholesky decomposition and its algorithm are presented, 
a corresponding routine in C is given 
and comments follow 
before concluding. 

\section{Reduction of the RPA problem using Cholesky decomposition} 
\label{Sprev} 

The eigenvalue problem of eq.~(\ref{Erpa}) leads in a straightforward manner 
to the system of equations 
\begin{equation} 
\left\{ 
\begin{array}{l} 
(A-B)(X_{\lambda}-Y_{\lambda}) = \varepsilon_{\lambda} (X_{\lambda}+Y_{\lambda}) \\ 
(A+B)(X_{\lambda}+Y_{\lambda}) = \varepsilon_{\lambda} (X_{\lambda}-Y_{\lambda}) 
\end{array} 
\right\} 
\label{Erpasys1}  
\end{equation} 
or   
\begin{equation} 
\left\{ 
\begin{array}{l} 
(A+B)(A-B)(X_{\lambda}-Y_{\lambda}) = \varepsilon_{\lambda}^2(X_{\lambda}-Y_{\lambda}) \\ 
(A-B)(A+B)(X_{\lambda}+Y_{\lambda}) = \varepsilon_{\lambda}^2(X_{\lambda}+Y_{\lambda}) 
\end{array} 
\right\} .  
\label{Erpasys}  
\end{equation} 
The matrices $A\pm B$ are real and symmetric. 
Let us assume that $A+B$ is positive-definite. 
Then it can be factorized as 
\begin{equation}
 A+B = LL^T 
, 
\end{equation} 
where $L$ a lower-triangular real matrix and $L^T$ its transpose. 
This is the square-root or Cholesky decomposition of the matrix. 
Then the second equation of the system (\ref{Erpasys}), 
premultiplied with $L^T$,  
can be written as 
\begin{equation} 
HR_{\lambda} = \varepsilon_{\lambda}^2R_{\lambda}     , 
\label{Erpach} 
\end{equation} 
where 
\begin{equation}
H \equiv L^T(A-B)L 
\end{equation} 
a real symmetric matrix 
and 
\begin{equation}
R_{\lambda} \equiv  
\varepsilon_{\lambda}^{-1/2}L^T(X_{\lambda}+Y_{\lambda}) 
\end{equation} 
its orthonormalized eigenvectors with eigenvalues $\varepsilon_{\lambda}^2$. 

The original RPA problem has been reduced to a symmetric problem of half the dimension. 
For $\varepsilon_{\lambda}^2>0$ the eigenvalues of the original problem 
are real, $\varepsilon_{\lambda}=\pm\sqrt{\varepsilon_{\lambda}^2}$.  
Once the eigenvectors $R_{\lambda}$ have been evaluated, the vectors $X_{\lambda}$ and $Y_{\lambda}$ can be recovered. 
For real and positive eigenvalues (and real eigenvectors), 
\begin{eqnarray} 
X_{\lambda} &=& \frac{1}{2} [\sqrt{\varepsilon_{\lambda}}(L^T)^{-1} + \frac{1}{\sqrt{\varepsilon_{\lambda}}}L]R_{\lambda}, \\  
Y_{\lambda} &=& \frac{1}{2} [\sqrt{\varepsilon_{\lambda}}(L^T)^{-1} - \frac{1}{\sqrt{\varepsilon_{\lambda}}}L]R_{\lambda}.  
\end{eqnarray} 
It is easily verified that the desired normalization condition 
as well as the orthogonality condition between different eigenvectors 
\begin{equation} 
X_{\lambda}^T X_{\mu} - Y_{\lambda}^T Y_{\mu} = \delta_{\lambda\mu} 
\label{Exyorth} 
\end{equation} 
are satisfied.  

If the matrix $A+B$ is not positive definite, but $A-B$ is, 
one can write 
$ 
 A-B = LL^T 
$
and proceed in an analogous way by 
rewriting the first equation of the system (\ref{Erpasys}) 
in the form (\ref{Erpach}) 
with 
\begin{equation} 
H \equiv L^T(A+B)L \quad , \quad  
R_{\lambda} \equiv  
\varepsilon_{\lambda}^{-1/2}L^T(X_{\lambda}-Y_{\lambda}) 
. 
\end{equation} 
From $R_{\lambda}$ the vectors $X_{\lambda}$ and $Y_{\lambda}$ can be recovered, 
\begin{equation} 
X_{\lambda} = \frac{1}{2} [\frac{1}{\sqrt{\varepsilon_{\lambda}}}L +\sqrt{\varepsilon_{\lambda}}(L^T)^{-1}]R_{\lambda} 
, \end{equation} 
\begin{equation} 
Y_{\lambda} = \frac{1}{2} [\frac{1}{\sqrt{\varepsilon_{\lambda}}}L -\sqrt{\varepsilon_{\lambda}}(L^T)^{-1}]R_{\lambda} 
. 
\end{equation} 
For more details see Ref.~\cite{Chi1970}. 

Among the virtues of this method of solving the RPA problem are that 
the Cholesky decomposition is fast and efficient numerically \cite{PTV1997} 
and that one only has to deal with triangular and symmetric matrices, 
which simplifies the numerical realization of the solution. 

\section{Generalized methods} 
\label{Sour} 

Unfortunately, if neither $A+B$ nor $A-B$ is positive definite, 
the above method fails and one has to either use general routines to 
solve the full non-symmetric $(2n)\times (2n)$ RPA problem 
or use complex $n \times n$ matrices (see Ref.~\cite{UlR1971}; 
alternatively, one can define a Cholesky decomposition $LL^T$ where imaginary values 
are allowed in $L$~\cite{Fad1959}).  

One can still reduce the size of the problem using other decompositions, 
valid for generic real symmetric matrices 
\cite{Fad1959,Fox1964,GoV1996} or, in fact, no decomposition at all, as will be shown. 
For the moment, the discussion is kept general. 
Let us consider the second equation of the system (\ref{Erpasys}). 
One can of course start with the first one and proceed in a completely 
analogous way. 
Suppose we can factorize the matrix $(A+B)$ as 
\begin{equation}  
A+B = CDE ,  
\end{equation}  
where $C$, $D$, $E$ are $n \times n$ matrices 
and $E$ has an inverse. 
(Any one of them may be the identity matrix $I$). 
The equation in question can be written as 
\begin{equation} 
HR_{\lambda} = \varepsilon_{\lambda}^2R_{\lambda} 
\label{EHRgen} 
\end{equation} 
with 
\begin{equation} 
H\equiv E(A-B)CD \quad , \quad R_{\lambda} = \frac{1}{\sqrt{\varepsilon_{\lambda}}} E(X_{\lambda}+Y_{\lambda}) 
. 
\label{Hgen} 
\end{equation} 
Henceforth we consider the solutions with real and positive $\varepsilon_{\lambda}$. 
Real eigenvalues allow us to assume real vectors $R_{\lambda}$.  
The real vectors 
\begin{eqnarray} 
X_{\lambda} &=& \frac{1}{2}[{\sqrt{\varepsilon_{\lambda}}} E^{-1} + \frac{1}{\sqrt{\varepsilon_{\lambda}}}CD]R_{\lambda}   
, \label{EXgen} \\  
Y_{\lambda} &=& \frac{1}{2}[{\sqrt{\varepsilon_{\lambda}}} E^{-1} - \frac{1}{\sqrt{\varepsilon_{\lambda}}}CD]R_{\lambda} 
\label{EYgen} 
\end{eqnarray} 
form solutions of the RPA problem, eq.~(\ref{Erpa}), as can be easily verified. They 
obey the orthonormalization condition 
\begin{equation} 
X_{\lambda}^TX_{\mu} - Y_{\lambda}^T Y_{\mu} = \pm \delta_{\lambda\mu} 
\label{Exynewnorm} 
, 
\end{equation} 
which is equivalent to 
\begin{equation} 
R_{\lambda}^T   (E^{-1})^TCD R_{\mu} 
= \pm \delta_{\lambda\mu}  
. 
\label{Rgen} 
\end{equation} 
Note that 
\begin{equation} 
(E^{-1})^TCD  
=D^T C^T E^{-1} =(E^{-1})^T(A+B)E^{-1}. 
\end{equation} 
Expressions (\ref{EXgen}), (\ref{EYgen}) can be obtained by writing 
$\{X_{\lambda} \, \mbox{or} \, Y_{\lambda}\} = \frac{1}{2}[(X_{\lambda}+Y_{\lambda}) \{+\,\mbox{or}\,-\} (X_{\lambda}-Y_{\lambda})]$, 
then writing $(X_{\lambda}-Y_{\lambda})=\varepsilon_{\lambda}^{-1}(A+B)(X_{\lambda}+Y_{\lambda})$ 
(see eq.~(\ref{Erpasys1})) 
and finally using $R_{\lambda}=\sqrt{\varepsilon_{\lambda}}E^{-1}(X_{\lambda}+Y_{\lambda})$ and the decomposition of $(A+B)$. 
Eq.~(\ref{Rgen}) is easily obtained if one substitutes $X$ and $Y$ in 
eq.~(\ref{Exynewnorm}) 
with expressions (\ref{EXgen}) and (\ref{EYgen}).  
Thus, the existence of the orthonormalizable real solutions $(X_{\lambda,\mu},Y_{\lambda,\mu})$ 
for the RPA matrix 
and of the inverse of $E$  
guarantee the orthogonality condition eq.~(\ref{Rgen}) for the corresponding eigenvectors of $H$  
(which has been verified by means of numerical examples for the special cases presented in the 
next two subsections). 

The normalization condition does not hold automatically. 
Therefore, either the eigenvectors 
$R_{\lambda}$ have to be renormalized according to eq.~(\ref{Rgen}) (for $\lambda = \mu$) 
after solving the eigenvalue problem (\ref{EHRgen}), 
or the $X_{\lambda}$ and $Y_{\lambda}$ vectors have to be renormalized 
according to eq.~(\ref{Exynewnorm}), after they are calculated from $R_{\lambda}$ and 
eqs.~(\ref{EXgen}), (\ref{EYgen}). 
Regarding the r.h.s. of eq.~(\ref{Exynewnorm}), note that 
the norm of the 
RPA eigenvectors -- as defined by the l.h.s. --  
need not be positive, unless the 
stability matrix is positive definite~\cite{Tho1961}. 
If it is not, eigenvectors with positive eigenvalues may exist which are 
normalizable to $-1$, instead of $1$. 
Such a case can be recognized before normalization 
by evaluating the l.h.s. of expression (\ref{Exynewnorm}) or (\ref{Rgen}) 
and checking the sign of the result. 

In Ref.~\cite{UlR1971} an orthogonal transformation was utilized, $A\pm B = CDC^T$, 
where $C$ orthogonal and $D$ diagonal. 
(It was applied in a different way 
so as to obtain a symmetric problem -- real or complex.)  
In the method described previously 
based on the usual Cholesky decomposition, 
one has $C=L$, $D=I$ and $E=L^T$ 
and the generic equations (\ref{Hgen})--(\ref{Rgen}) 
simplify into the ones presented earlier 
and in Ref.~\cite{Chi1970}. 
Indeed, a procedure such as outlined above 
would make less sense to apply if the matrices $C$, $D$, $E$ did not have special properties 
which simplify the algebra and numerics involved.  

\subsection{No decomposition} 

One may chose simply to solve one of the $n \times n$ eigenvalue problems 
of eq.~(\ref{Erpasys}) without any decomposition. 
For example, one can solve the second one 
for $R_{\lambda}=\varepsilon_{\lambda}^{-1/2}(X_{\lambda}+Y_{\lambda})$ 
and apply the equations given above for 
\begin{equation} 
D=E=I \quad , \quad C=A+B. 
\end{equation} 
Some matrix operations are thus saved, 
but the resulting eigenvalue problem is not symmetric. 
The no-decomposition strategy should be preferable when 
the matrices are expected to not be positive-definite.  

\subsection{Generalized Cholesky decomposition} 

A decomposition strategy like the one discussed next 
offers the additional possibility to detect and solve a real-symmetric problem, 
if the decomposed matrix turns out to be positive-definite. 
It can be shown that a non-singular real symmetric matrix $F$ 
with non-singular leading submatrices, 
with or without negative eigenvalues, can be factorized in the form 
\begin{equation} 
F = L D L^T , 
\label{Echgen}  
\end{equation} 
where $L$ is again a lower-triangular matrix and $D=\mathrm{diag}\{d_i\}$ is a diagonal matrix whose 
diagonal elements are equal to $1$ or $-1$.
Indeed, the factorization (\ref{Echgen}) 
follows from the 
``$LDL$" decomposition of linear algebra, let us write it as 
\begin{equation} 
F=L'D'{L'}^T 
, \label{Eldl} \end{equation} 
where $L'$ is a unit-triangular matrix and $D'=\mathrm{diag}\{d_i'\}$ 
a general diagonal matrix~\cite{Fox1964,GoV1996}.  
If we write $D'$ as 
$$ D'= [\mathrm{diag}\{\sqrt{|d'_i|}\}] \, 
[\mathrm{diag}\{\mathrm{sgn} d'_i\}]  \,
[\mathrm{diag}\{\sqrt{|d'_i|}\}]  
$$ 
and define $L$ and $D$ as  
$$ 
L = L' \, [\mathrm{diag}\{\sqrt{|d'_i|}\}] 
\,\, 
, 
\,\, 
D =  
\mathrm{diag}\{\mathrm{sgn}d'_i\}  
, 
$$  
eq.~(\ref{Echgen}) is obtained. 
The existence of the ``$LDL$" decomposition (eq.~(\ref{Eldl}))  
for a non-singular real symmetric matrix $F$ with non-singular leading submatrices 
is known from linear algebra~\cite{Fox1964,GoV1996}. 

Setting $C=L$ and $E=L^T$ in 
eqs.~(\ref{Hgen})--(\ref{Rgen}) 
we obtain the $n\times n$ real, non-symmetric (in general) eigenvalue problem 
of the form (\ref{EHRgen}) 
with 
\begin{equation} 
H\equiv L^T(A-B)LD \quad , \quad R_{\lambda} = \frac{1}{\sqrt{\varepsilon_{\lambda}}} L^T(X_{\lambda}+Y_{\lambda}) 
. 
\label{Hnew} 
\end{equation} 
At this point one may check whether $D=I$ (meaning that $F$ is positive-definite), 
in which case $H$ is symmetric and 
optimized routines can be used to solve its eigenvalue problem. 

Applying eqs.~(\ref{EXgen}) (\ref{EYgen}) we find that 
for real and positive eigenvalues  
the solutions of the RPA equations are given by 
the vectors 
\begin{eqnarray} 
X_{\lambda} &=& \frac{1}{2}[{\sqrt{\varepsilon_{\lambda}}} (L^T)^{-1} + \frac{1}{\sqrt{\varepsilon_{\lambda}}}LD]R_{\lambda},  \\ 
Y_{\lambda} &=& \frac{1}{2}[{\sqrt{\varepsilon_{\lambda}}} (L^T)^{-1} - \frac{1}{\sqrt{\varepsilon_{\lambda}}}LD]R_{\lambda} 
.\end{eqnarray} 
The normalization condition (\ref{Rgen}) is simplified since 
$(E^{-1})^TCD = D$. 
 

The ``generalized" Cholesky decomposition defined by eq.~(\ref{Echgen}) 
can be realized numerically almost as efficiently as the usual one. 
A simple code in C is given later on. 
The underlying algorithm 
(a revision of a similar one which appears in Ref.~\cite{Two1996}) 
resembles closely the usual Cholesky algorithm and can be described as follows. 

If we write out eq.~(\ref{Echgen}) in components 
($F=[f_{ij}]$, $D=[d_{ij}]$ with $d_{ij}=d_{i}\delta_{ij}$, $L=[l_{ij}]$), 
we have 
\begin{equation} 
f_{ij} = \sum_{k\leq i,j} d_k l_{ik} l_{jk} ,  
\end{equation} 
from which we readily obtain, setting $i=j$ and $i<j$ respectively,  
\begin{equation} 
l_{ii} = \sqrt{d_i (f_{ii} - \sum_{k < i } d_k l_{ik}^2) } 
, \end{equation} 
\begin{equation} 
l_{ji} = \frac{d_i}{l_{ii}} (f_{ji} - \sum_{k < i } d_k l_{ik}l_{jk})  . 
\label{Ealg} 
\end{equation} 
We start the solution with $l_{11}=\sqrt{d_1f_{11}}=\sqrt{|f_{11}|}$; 
if $f_{11}<0$, we have $d_1=-1$, 
otherwise $d_1=1$. 
The off-diagonal elements $l_{j1}$ can now be evaluated. 
After finishing with the first column of $L$, 
we continue with $l_{22}$. Similarly, the sign of $d_2$ will be 
determined by the sign of $(f_{22} - d_1 l_{11}^2)$. 
All quantities needed to calculate $l_{j2}$ are now known. 
In short, we apply eqs.~(\ref{Ealg}) in the order $i=1,2,\ldots,n$, 
each time starting with the diagonal element $l_{ii}$ and proceeding with 
the off-diagonal elements in the same column. 

\section{A routine in C} 

A simple routine example is given below in C language, 
based on the 
{\tt choldc} 
routine of \cite{PTV1997}, \S~2.9 
and the algorithm described previously. 
Given the real-symmetric $n \times n$ matrix $F$  
({\tt f[1...n][1...n]}), 
this routine constructs its decomposition  
$F=L D L^T$, where $D$ is a diagonal matrix with 
elements equal to 1 or $-1$ along the diagonal and $L$ a lower-triangular matrix. 
Only components $F_{ij}$ with $j\geq i$ need to be referenced, since $F$ is symmetric. 
This allows the elements of $L$ lying below the diagonal to be stored in the 
corresponding elements of {\tt f}. 
Additional arrays 
{\tt p[1...n]} and {\tt d[1...n]}  
are used to store the diagonals of $L$ 
and $D$ respectively.
One can also use the routine to test if the matrix $F$ is positive-definite, 
by looking for $-1$ elements in $D$. 
The return value can, {\em e.g.},  be an integer (instead of void) equal to the number of negative 
{\tt d[i]} values. 
For matrices which are expected to have mostly positive eigenvalues, in which 
case most {\tt d[i]}'s will be equal to $+1$, 
it should be advantageous to only store the {\tt i}-values for which 
{\tt d[i]}$=-1$, instead of the whole array {\tt{d}}, 
especially if the matrix dimension is large. 

\begin{verbatim}  
...
float f[N][N];  // N=n+1   
float p[N]; 
int   d[N]; 
...
void choldmod(int n) 
{ 
  int i,j,k; 
  float sum,aux; 

  for (i=1;i<=n;i++) 
  { 
    d[i]=1;  
    for (j=i;j<=n;j++) 
    { 
      sum=0.0;  
      for(k=1; k<i; k++) 
          sum += d[k]*f[i][k]*f[j][k]; 
      aux = f[i][j]-sum; 
      if (i==j) 
      { 
         if (aux<=0.0) d[i]=-1; 
         p[i]=sqrt(d[i]*aux); 
      } 
      else 
         f[j][i]=d[i]*aux/p[i];  
    }
  }  
} 
\end{verbatim}

The inner loop, consisting of two multiplies (one of them with {\tt int}) and a sum, 
is executed $(n^3-n)/6 \approx n^3/6$ times. There are also $n$ square roots and about $n^2/2$ divides. 
On an Intel Pentium 4 machine ({\tt gcc} compilation) an arbitrary 
$1000 \times 1000$ symmetric matrix was factorized in 
about 2 seconds and a $10000 \times 10000$ matrix in about 40 minutes. 
The algorithm of the usual Cholesky decomposition contains $n(n^2-1)/6\approx n^3/6$  
similar inner loops (with one multiply and one subtract each), 
$n$ square roots and about $n^2/2$ divides~\cite{PTV1997}. 
A similar amount of effort is required for a usual $LDL$ decomposition, 
whereas 
the orthogonal transformation proceeding via a Householder reduction 
and a $QL$ decomposition requires more effort, 
namely about $2n^3/3 + 30n^2$ operations~\cite{PTV1997}. 
As holds for the usual $LDL$ factorization, 
stability of the present 
algorithm is not guaranteed for indefinite matrices. 
Since, however, RPA matrices tend to have their largest elements 
(absolute values) along the diagonal, they should be mostly well behaved. 

\section{Comments} 

In most of the above it is assumed that there are no singularities. 
In actual applications it is quite improbable 
to obtain eigenvalues so close to zero that bad behaviour occurs. 
Even then, 
one can escape the pitfall by 
slightly scaling the 
residual interaction entering the RPA equation and repeating the calculation. 

The modified eigenvalue problem defined by eqs.~(\ref{EHRgen}) and (\ref{Hnew}) 
can be transformed into a complex symmetric problem. 
One has to multiply from the left both sides of eq.~(\ref{EHRgen}) with the 
square-root matrix of $D$, {\em i.e.}, with 
the diagonal matrix $\tilde{D}$ with diagonal elements $\tilde{d}_i = \sqrt{d_i}$ equal to 
1 or i, so that $D=\tilde{D}^2$. 
The $i-$th element of the eigenvector of the new, symmetric (but complex, if $D$ contains negative elements) 
matrix $\tilde{D}L^T(A-B)L\tilde{D}$,  
is equal to the $i-$th element of the corresponding eigenvector of $H$, 
times the $i-$th diagonal element of $\tilde{D}$. 
That the RPA problem can be reduced in a complex symmetric problem 
was shown in a different way in Ref.~\cite{UlR1971}. 

The modified Cholesky decomposition defined here is not available in packages like LAPACK, 
appropriate for very large matrices, 
contrary to the usual Cholesky and other decompositions. 
Therefore, for very large matrices one may have to chose another reduction procedure.

\section{Conclusion} 
\label{Sconc} 

In conclusion, it has been demonstrated that there exist methods 
to reduce a real RPA eigenvalue problem to a 
real non-symmetric problem of half the dimension without demanding that 
one of the matrices $A+B$, $A-B$ be positive definite. 
Reduction can be achieved with or without matrix decomposition. 
We have worked out a method based on a generalized Cholesky decomposition,
which is no more involved numerically than the one relying on the 
usual square-root decomposition of positive-definite real-symmetric matrices. 
The result is in general a real non-symmetric (or a complex-symmetric) eigenvalue problem of half 
the dimension. If one of the matrices turns out to be positive definite, a real-symmetric problem is 
obtained instead, without additional effort.  

\acknowledgments
Work supported by the Deutsche Forschungsgemeinschaft, contract SFB634.

\end{document}